\documentclass[aps,prl,onecolumn,notitlepage,citeautoscript,superscriptaddress,footinbib,longbibliography,11pt]{revtex4-2}
\usepackage{times}
\usepackage{amsmath,amssymb,amsfonts,mathrsfs,bm,feynmf,setspace}
\usepackage{dcolumn}
\usepackage{graphicx}
\usepackage{latexsym}
\usepackage{epsfig}
\usepackage{floatrow}
\usepackage{ulem}

\begin{document}

\title{Ferroelectricity-driven phonon Berry curvature and non-linear phonon Hall transports}

\author{Jino \surname{Im}}

\affiliation{Chemical Data-driven Research Center, Korea Research Institute of Chemical Technology, Daejeon 34114, Korea}

\author{Choong~H. \surname{Kim}}
\email[Corresponding author.\\]{chkim82@snu.ac.kr}

\affiliation{Center for Correlated Electron Systems, Institute for Basic Science (IBS), Seoul 08826, Korea}
\affiliation{Department of Physics and Astronomy, Seoul National University, Seoul 08826, Korea}

\author{Hosub \surname{Jin}}
\email[Corresponding author.\\]{hsjin@unist.ac.kr}
\affiliation{Department of Physics, Ulsan National Institute of Science and Technology (UNIST), Ulsan 44919, Korea}

\begin{abstract}
Berry curvature (BC) governs topological phases of matter and generates anomalous transport. When a magnetic field is applied, phonons can acquire BC indirectly through spin-lattice coupling, leading to a linear phonon Hall effect. Here, we show that polar lattice distortion directly couples to a phonon BC dipole, which causes a switchable non-linear phonon Hall effect. In a SnS monolayer, the in-plane ferroelectricity induces a phonon BC and leads to the phononic version of the non-volatile BC memory effect. As a new type of ferroelectricity-phonon coupling, the phonon Rashba effect emerges and opens a mass-gap in tilted Weyl phonon modes, resulting in a large phonon BC dipole. Furthermore, our \textit{ab initio} non-equilibrium molecular dynamics simulations reveal that non-linear phonon Hall transport occurs in a controllable manner via ferroelectric switching. The ferroelectricity-driven phonon BC and corresponding non-linear phonon transports provide a novel scheme for constructing topological phononic transport/memory devices.
\end{abstract}  



\begin{center}
  {\large\bf Ferroelectricity-driven phonon Berry curvature and non-linear phonon Hall transports}\\\vspace{0.1cm}
  Jino Im$^{1,\dagger}$, Choong H. Kim$^{2,3,\dagger,*}$, and Hosub Jin$^{4,*}$\\
  {\it\small
    $^1$Chemical Data-driven Research Center, Korea Research Institute of Chemical Technology, Daejeon 34114, Korea\\
    $^2$Center for Correlated Electron Systems, Institute for Basic Science (IBS), Seoul 08826, Korea\\
    $^3$Department of Physics and Astronomy, Seoul National University, Seoul 08826, Korea\\
    $^4$Department of Physics, Ulsan National Institute of Science and Technology (UNIST), Ulsan 44919, Korea
    }
\end{center}
\rule{8cm}{0.01cm}

\noindent
{\small
$^{\dagger}$These authors contributed equally to this work.\\
$^*$Corresponding author. Email: chkim82@snu.ac.kr (C.H.K.); hsjin@unist.ac.kr (H.J.)\\
}


\noindent
Berry curvature (BC) governs topological phases of matter and generates anomalous transport. When a magnetic field is applied, phonons can acquire BC indirectly through spin-lattice coupling, leading to a linear phonon Hall effect. Here, we show that polar lattice distortion directly couples to a phonon BC dipole, which causes a switchable non-linear phonon Hall effect. In a SnS monolayer, the in-plane ferroelectricity induces a phonon BC and leads to the phononic version of the non-volatile BC memory effect. As a new type of ferroelectricity-phonon coupling, the phonon Rashba effect emerges and opens a mass-gap in tilted Weyl phonon modes, resulting in a large phonon BC dipole. Furthermore, our \textit{ab initio} non-equilibrium molecular dynamics simulations reveal that non-linear phonon Hall transport occurs in a controllable manner via ferroelectric switching. The ferroelectricity-driven phonon BC and corresponding non-linear phonon transports provide a novel scheme for constructing topological phononic transport/memory devices.

\textbf{\large Introduction}

In a symmetry-broken environment, the Bloch electrons of a crystalline system can acquire Berry curvature (BC), which leads to a non-trivial band topology or anomalous charge transport~\cite{PhysRevLett.49.405,RevModPhys.82.1959}. For example, in broken time-reversal symmetry, the electronic structure acquires a finite monopole component of the BC, referred to as a Berry phase, which induces the linear anomalous Hall effect~\cite{RevModPhys.82.1539}. On the other hand, breaking the inversion symmetry allows the dipole component of the BC to survive and provides a non-linear anomalous Hall effect and circular photo-galvanic effect, even with preservation of the time-reversal symmetry~\cite{arxiv0904.1917,PhysRevLett.105.026805,PhysRevLett.115.216806,Xu2018,Ma2019,Orenstein2021}.
\vspace{2mm}

Phonons, which are bosonic quasiparticles composed of collective lattice vibrations, can also host BC through symmetry breaking. Contrary to the common idea about the charge neutrality of phonons, a time-reversal breaking magnetic field can couple with them, giving rise to the phonon BC and, consequently, the phonon Hall effect~\cite{PhysRevLett.95.155901,PhysRevLett.96.155901,PhysRevLett.100.145902,PhysRevLett.105.225901}. Phonons with non-zero BC traverse perpendicular to both the magnetic field and temperature gradient. In general, however, due to indirect coupling through weak spin-phonon interactions~\cite{Kronig1939,PhysRev.57.426,PhysRev.119.1204}, the magnetic field cannot efficiently create phonon BC, making the phonon Hall effect less practical. If we turn our focus to another symmetry-breaking factor, breaking the inversion symmetry could be more effective for creating the phonon BC, as phonons are inherently composed of lattice degrees of freedom and are directly coupled to inversion-breaking lattice distortions. In an inversion-broken environment, therefore, a large phonon BC can emerge and govern unprecedented phonon transports, such as the phonon analogues of the non-linear Hall and circular photo-galvanic effects.
\vspace{2mm}

\begin{figure}[t]
  \includegraphics[width=0.65\textwidth]{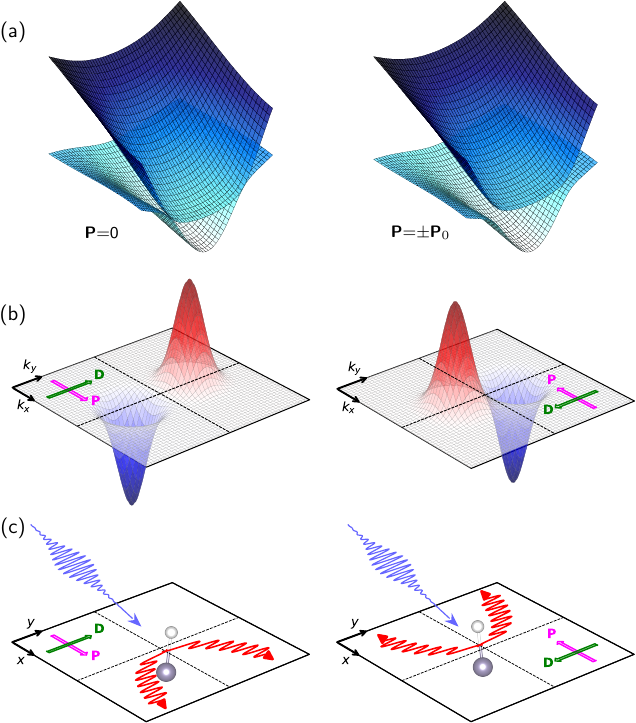}
  \caption{
    Phonon Berry curvature (BC) dipole and non-linear phonon Hall effect induced by ferroelectricity.
    (a) Tilted Weyl phonons undergo gap opening by ferroelectric polarization.
    (b) Phonon BC dipole develops perpendicular to the polarization and reverses its direction by ferroelectric switching.
    (c) Non-linear phonon Hall current through the ferroelectricity-induced phonon BC dipole.
  }
\end{figure}

Here, we show that ferroelectric polarization, serving as an inversion-breaking order parameter, can generate a large phonon BC and its dipole component. In a SnS monolayer, a tilted Weyl phonon is formed at the band crossing between the longitudinal and transverse optical phonon modes, and undergoes gap-opening by in-plane ferroelectric polarization (Fig.~1(a)). As a new type of lattice-phonon interaction, the phonon Rashba effect is caused by ferroelectric displacement, and represents the microscopic origin of the mass-gap in the tilted Weyl phonon. The ferroelectricity-driven phonon Rashba effect is also characterized by a unique chiral phonon texture, and the chirality can be switched by ferroelectric reversal. As a result, a large phonon BC dipole arises from the tilted and gapped Weyl phonon, which can be controlled by ferroelectric switching (Fig.~1(b)). Analogous to the relationship between the BC dipole and non-linear Hall effect in Bloch electrons, the phonon BC dipole represents a response function to the non-linear phonon Hall effect. Our {\it ab initio} non-equilibrium molecular dynamics simulations showed that transverse phonon currents arise in response to linear and circular vibrating atoms, demonstrating the non-linear phonon Hall effect (Fig.~1(c)).
\vspace{4mm}

\textbf{\large Results}

\textbf{Ferroelectricity-induced phonon Berry curvature in a SnS monolayer}

\begin{figure}[t]
 \centerline{\includegraphics[width=0.7\textwidth]{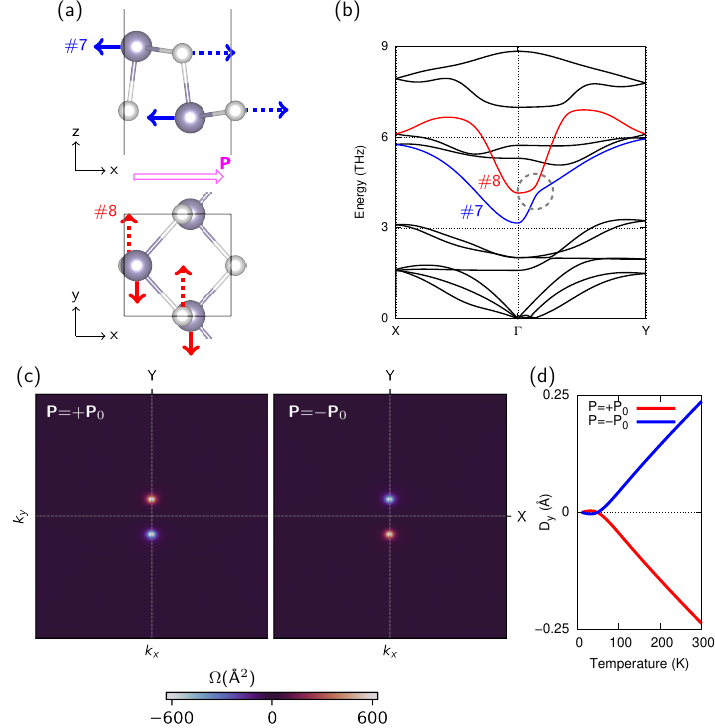}}
 \caption{
  A tilted and gapped Weyl phonon in a SnS monolayer.
  (a) Side view and top view of the SnS monolayer. The normal modes of the \#7 and \#8 branches are indicated by blue and red arrows, respectively. The solid and dashed arrows represent the displacement of Sn and S atoms, respectively.
  (b) Phonon dispersion of the SnS monolayer.
  (c) BC distributions of the \#7 phonon band under the positive and negative polarization directions.
  (d) Temperature dependence of phonon BC dipole.
 }
\end{figure}

\indent A SnS monolayer with {\it Pnm2$_1$} symmetry has in-plane ferroelectric polarization, as shown in Fig.~2(a), where Sn and S atoms undergo opposite displacements along the [100] direction~\cite{Higashitarumizu2020,Wu2016,PhysRevLett.117.097601}. The system has mirror reflection symmetry with the $xz$ mirror plane on which the ferroelectric polarization resides. The mirror symmetry combined with inversion asymmetry allows for a non-zero electronic BC dipole perpendicular to the mirror plane in two-dimensional systems~\cite{PhysRevLett.115.216806}. From the ferroelectric-driven BC dipole, non-linear Hall and circular photo-galvanic effects controlled by ferroelectric switching arise~\cite{Kim2019,Wang2019}.
\vspace{2mm}

From first-principles phonon calculations (see Supporting Information), we identified a gapped and tilted Weyl cone consisting of the 7$^{th}$ longitudinal and 8$^{th}$ transverse optical branches (denoted as \#7 and \#8 bands, respectively), as shown in Fig.~2(b). As illustrated by the arrows in Fig.~2(a), \#7 and \#8 modes at $\Gamma$ point exhibit the opposite displacement of Sn and S atoms along the $x$- and $y$-directions, respectively. As in electronic structures, a tilted Weyl cone with a small gap is a source of large BCs~\cite{PhysRevLett.115.216806,PhysRevLett.121.246403,PhysRevB.97.041101}. As the two Weyl cones located at ${\bf k}$~=~$\pm {\bf k}_0$~=~(0, $\pm$0.11 \AA$^{-1}$) are time-reversal partners, they have the opposite sign of phonon BC $(\Omega(+{\bf k}_0)=-\Omega (-{\bf k}_0))$, giving rise to a phonon BC dipole along the $y$-direction. The phonon BC of the \#7 phonon band is shown in Fig.~2(c). Large phonon BC peaks with maximum magnitude reaching $\sim$ 600 \AA$^2$ appear at $\bf{k}=\pm \bf{k}_0$. If the ferroelectric polarization is flipped, the sign of BC changes and the direction of the BC dipole is also reversed. As the two opposite ferroelectric configurations are inversion partners, the phonon BC satisfies $\Omega^{+{\bf P}}(+{\bf k})=\Omega^{-{\bf P}}(-{\bf k})$. When combined with $\Omega(+{\bf k})=-\Omega(-{\bf k})$ imposed by the time-reversal symmetry, we have $\Omega^{+{\bf P}}({\bf k})=-\Omega^{-{\bf P}}({\bf k})$. Figure~2(d) shows the temperature dependence of the phonon BC dipole pointing in the $y$-direction. It exhibits almost linear behaviour as a function of temperature, and its absolute value approaches 0.24 \AA~ at room temperature. The temperature dependence of the phonon BC dipole is calculated only through the Bose-Einstein distribution, and the polarization strength is fixed at a value of 0 K. This constant polarization approximation is plausible because the theoretically estimated $T_{\mathrm{c}}$ of the SnS monolayer is 1,200 K~\cite{PhysRevLett.117.097601}, which is much higher than room temperature.
At 300 K, the BC dipole from the \#7 phonon band is not shaded by other phonon modes that are randomly excited by temperature.
Due to inherent and direct coupling between ferroelectricity and the phonon, ferroelectric distortion becomes an effective tool for generating large values of the phonon BC and its dipole; this can cause non-linear transverse phonon transport, as described later.
\vspace{2mm}

\textbf{Phonon Rashba effect as a ferroelectricity-phonon interaction}
\\
\indent To understand the phonon BC distribution of the SnS monolayer, it is necessary to investigate how ferroelectricity develops a gap in the tilted Weyl phonon. The \#7 and \#8 phonon bands are shown in three different configurations in Fig.~3(a): two ferroelectric configurations with opposite polarizations (${\bf P}$~=~$\pm {\bf P}_0$) and one paraelectric case (${\bf P}$~=~${\bf 0}$). Without the ferroelectric distortion (${\bf P}$~=~${\bf 0}$), the \#7 and \#8 bands cross each other to form gapless and tilted Weyl cones along the $\Gamma$-Y direction. Once the ferroelectric polarization is turned on (${\bf P}$~=~$\pm {\bf P}_0$), the Weyl cones are gapped, clearly demonstrating that ferroelectricity forms a gap in the tilted Weyl phonons. At the same time, the ferroelectric polarization imprints chirality at each Weyl cone, denoted by pseudo-angular momentum ($\bf{L}$) textures~\cite{PhysRevLett.115.115502,Zhu2018,PhysRevLett.121.175301}. By calculating the pseudo-angular momentum in the phonon dispersion, a finite expectation value of $L_z \equiv \left(\vert u_+ \rangle \langle u_+ \vert - \vert u_- \rangle \langle u_- \vert \right)=-i\left(\vert u_x \rangle \langle u_y \vert - \vert u_y \rangle \langle u_x \vert \right)$ appears only when the system has ferroelectricity. Here, $\vert u_{\pm} \rangle =\frac{1}{\sqrt{2}} \left(\vert u_x \rangle \pm i \vert u_y \rangle \right)$. Two Weyl phonons located at opposite points of the Brillouin zone (${\bf k}$~=~$\pm {\bf k}_0$) are the time-reversal partners; they have opposite chirality, which is linked to the different signs of BC. The chirality also depends on the direction of ferroelectric polarization. When the polarization is flipped, the chirality is reversed; this leads to reversal of the BC. In the SnS monolayer, ferroelectricity opens a gap at the tilted Weyl phonon and determines the chirality and BC textures, thus confirming the ferroelectricity-induced and -controlled phonon BC.
\vspace{2mm}

\begin{figure}[t]
 \centerline{\includegraphics[width=0.7\textwidth]{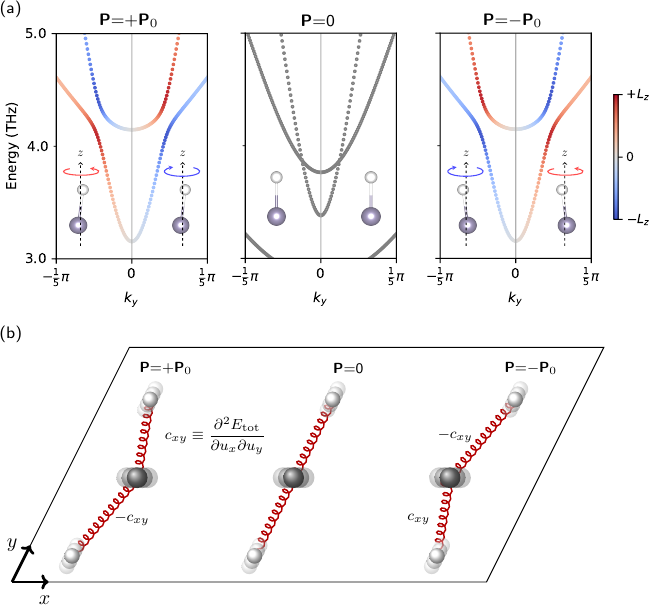}}
 \caption{
   Ferroelectricity-induced phonon Rashba effect and chiral texture.
  (a) The \#7 and \#8 phonon bands overlaid with the pseudo-angular momentum texture in three configurations (${\bf P}=\pm {\bf P}_0$ and ${\bf P}=0$).
  (b) The emergence of an anti-symmetric off-diagonal force constant ($c_{xy}$) induced by ferroelectric polarization.
}
\end{figure}

Figure 3(b) schematically shows the microscopic origin of the ferroelectricity-phonon interaction that triggers the ferroelectricity-induced chiral phonon texture and phonon BC in the SnS monolayer. The in-plane ferroelectric displacement allows two orthogonal vibration modes corresponding to the \#7 and \#8 phonon bands to intertwine. With the ferroelectric displacement, the off-diagonal element of the force constant matrix between the nearest $i^{th}$ and $j^{th}$ atoms located perpendicular to the polarization direction $\left( \mathrm{denoted~by}~ c_{xy} \equiv \frac {\partial^2 E_{\mathrm{tot}}}{\partial u_x^i \partial u_y^j} \right)$ survives. Without polarization (${\bf P}$~=~${\bf 0}$), $c_{xy}$ vanishes due to mirror reflection symmetry about the $yz$-plane. In addition, the off-diagonal force constant is anti-symmetric, {\it i.e.} the force constant between the nearest neighbour atom pair on the opposite side has the opposite sign $\left( \frac {\partial^2 E_{\mathrm{tot}}}{\partial u_x^i \partial u_y^j}=-\frac {\partial^2 E_{\mathrm{tot}}}{\partial u_x^j \partial u_y^i} \right)$ and the overall sign of the off-diagonal force constant is changed by the ferroelectric reversal. As a result, the ferroelectricity-phonon interaction in the dynamical matrix is written as
\begin{equation}
-i(\hat{\bf{P}}\cdot \hat{\bf{x}}) c_{xy} \sin k_y \left(\vert u_x \rangle \langle u_y \vert - \vert u_y \rangle \langle u_x \vert \right) \approx (\hat{\bf{P}}\cdot \hat{\bf{x}}) c_{xy} k_y L_z \equiv \alpha \left( \hat{\bf{P}} \times \bf{k}\right) \cdot \bf{L}.
\end{equation}
By analogy to the orbital Rashba effect of the Bloch electrons~\cite{Kim2019,Park2011}, this term can be referred to as the {\it phonon Rashba effect}. According to Eq.~(1), the chirality of phonons depends on both the position of the crystal momentum ($\bf{k}$) and the direction of ferroelectric polarization ($\bf{P}$). (For the full off-diagonal force constant in the SnS monolayer, see Supporting Information.)
\vspace{2mm}

Combining the phonon Rashba effect with the preformed tilted Weyl phonons, the effective Hamiltonian describing the two phonon modes is written as
\begin{equation}
\mathcal{H}_{\mathrm{eff}}=s\left[ v_x k_x \sigma_x + v_0 (k_y-sk_y^0)\sigma_0 - v_y(k_y-sk_y^0)\sigma_z +m\sigma_y\right],
\end{equation}
where $k_y^0 (=0.11$ \AA$^{-1})$ and $s(=\pm 1)$ denote the positions of two Weyl cones, $v_x$, $v_0$, and $v_y$ are the slopes of the Weyl cones, and $m\left(=(\hat{\bf{P}}\cdot \hat{\bf{x}})\alpha k_y^0\right)$ represents the mass-gap (see Supporting Information for details). Here, $\sigma$'s are the 2$\times$2 Pauli matrices based on $\left(\vert u_x \rangle, \vert u_y \rangle \right)$ vibration modes. When the ferroelectric polarization is turned on, the phonon Rashba effect emerges, as does $m$; the Weyl phonon opens a gap. The chiral charge of each Weyl cone is determined by the signs of $s$ and $m$. That is, chirality depends on both the position of the Weyl cone $(s)$ and the direction of the polarization $(m)$. The two Weyl cones located at $(0,s k_y^0)$ have opposite chirality and thus the opposite sign of BC, as shown in Fig.~2(c). The chirality and BC texture are reversed by ferroelectric switching (Fig.~3(a)). Ferroelectricity produces a unique phonon BC distribution in the SnS monolayer through the phonon Rashba effect.
\vspace{2mm}

When estimated from the mass-gap of the Weyl cone, the strength of the phonon Rashba effect in the SnS monolayer is $\sim 10^{-1} \omega_{\mathrm{D}}$, where $\omega_{\mathrm{D}}$ is the characteristic Debye frequency. The spin-phonon interaction is another source of the phonon BC and linear phonon Hall effect; it is realized by applying the external magnetic field, and its magnitude is known to be $\sim 10^{-4} \omega_{\mathrm{D}}$ in the paramagnetic dielectrics Tb$_3$Ga$_5$O$_{12}$~\cite{PhysRevLett.95.155901,PhysRevLett.105.225901}. The strength of the ferroelectricity-phonon interaction is three orders of magnitude larger than that of the spin-phonon interaction. In contrast to the indirect coupling between the magnetic field and phonons via spin-phonon interactions, ferroelectric distortions directly couple to phonons, providing an efficient way to generate phonon BCs. Moreover, given the non-volatile characteristics of ferroelectricity, the ferroelectricity-induced phonon BC can be utilized as a phonon BC memory~\cite{Xiao2020}.
\vspace{8mm}

\textbf{\large Discussion}

The phonon BC dipole in the SnS monolayer gives rise to the non-linear phonon Hall effect. In non-equilibrium molecular dynamics (NEMD) simulations with two different types of forced oscillations, we observed unique signatures of the non-linear phonon Hall effect and circular dichroic phonon rectification effect in the SnS monolayer. As illustrated in Fig.~4(a), we performed a NEMD simulation enforcing linear oscillation of a single pair of vertically aligned Sn and S atoms, where the direction of forced oscillation is perpendicular to that of ferroelectric polarization.
Figure 4(b) shows the temporal profiles of the differences in kinetic energy of atoms located in opposite directions with the forced oscillator interposed between them, which represents net phonon flow. (see Supporting Information). Along the direction of the forced oscillation ($y$-direction), the net phonon current is negligible. On the other hand, the phonon flow in the $x$-direction increases gradually with oscillation, which represents transverse phonon flow. The transverse signal consists of both direct current (DC) and alternating current (AC) components, and the frequency of the AC component is twice that of the forced oscillation; {\it i.e.} the DC and AC components correspond to the non-linear Hall and second harmonic generation of the phonon flow, respectively. After $t = 1.2$~ps, the DC component reaches its maximum and decreases because the phonon current touches the boundary of the periodic supercell. Snapshot images of the real-space kinetic energy distribution in Fig.~4(c) clearly demonstrate asymmetric phonon flow along the transverse direction. For detailed phonon transport, see Supporting Movie. The forced vibration of the local Sn-S pair partially excites the phonon modes responsible for the large BC dipole, and the perturbed non-equilibrium phonon distributions have a non-vanishing Berry phase. As a result, the transverse phonon current can flow perpendicular to the forced oscillation.
\vspace{2mm}

\begin{figure}[t]
  \centering
  \includegraphics[width=\textwidth]{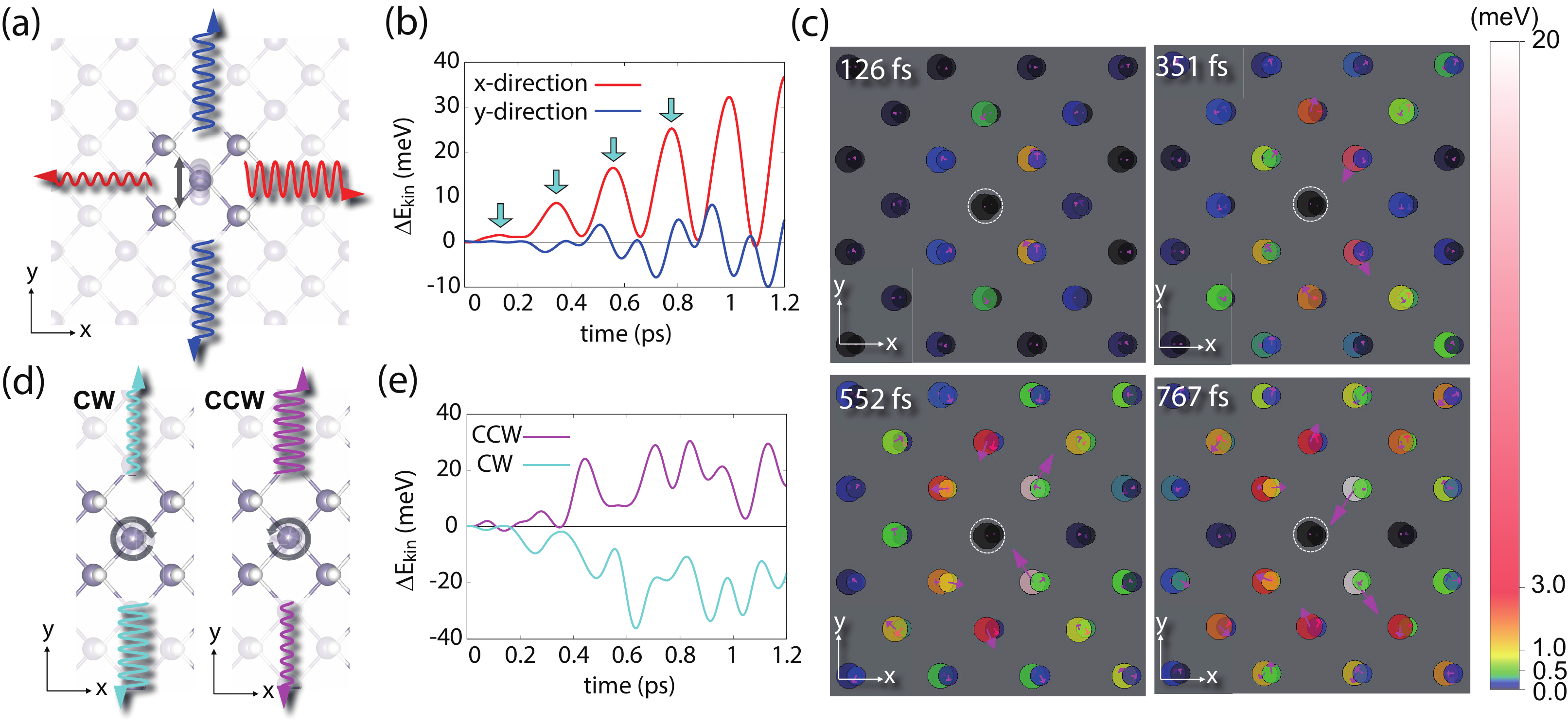}
  \caption{
    Switchable non-linear phonon Hall and circular dichroic phonon rectification effects.
    (a) Forced linear oscillation of a single pair of vertically aligned Sn and S atoms on the $y$-axis. Red and blue arrows represent phonon flow along the $x$-axis and $y$-axis, respectively.
    (b) Temporal profiles of kinetic energy differences along the $x$-axis (red) and $y$-axis (blue).
    (c) Snapshots of kinetic energy distribution in real space at 126 fs, 315 fs, 552 fs, and 767 fs after the initial oscillation. White dotted circles indicate a pair of the forced oscillation atoms.
    (d) Forced circular oscillations in clockwise (CW) and counterclockwise (CCW) directions. Magenta and cyan arrows represent phonon flow for CW and CCW circular oscillations, respectively.
    (e) Temporal profiles of kinetic energy differences along the $y$-axis for CW (magenta) and CCW (cyan) circular oscillations.
  }
\end{figure}

Another non-linear phonon response originating from the phonon BC can be captured by NEMD simulations when the system is driven by circularly forced oscillation of the single Sn-S pair. As shown in Fig.~4(d), we performed NEMD simulations with forced circular oscillations in the clockwise and counterclockwise directions. The temporal profiles of the kinetic energy difference showed net phonon flow along the $y$-direction; {\it i.e.} in response to circular perturbations, the phonon current is perpendicular to the ferroelectric polarization (Fig.~4(e)). The phonon transport generated by circular oscillations switches direction when the forced oscillation changes from clockwise to counterclockwise. This circular dichroic phonon rectification effect is a close analogue of the circular photo-galvanic effect originating from the electronic BC dipole~\cite{Xu2018,Kim2019}. In phonon rectification signals, the AC component has twice the frequency of the forced oscillation. When the ferroelectric polarization is flipped, the phonon current reverses its direction (see Supporting Information). Therefore, the circular dichroic phonon current can be readily switched by both the handedness of the circular perturbation and the ferroelectricity.
\vspace{2mm}

We estimated the fraction of the net transverse phonon transports normalized by the total flow propagating in all directions, which can be interpreted as a non-linear Hall angle. This ratio was about 30\% in the temporal range of 0.4 to 1.2 ps (see Supporting Information). A few tens of percent of the total phonon flow generated by the forced oscillation transformed into the net transverse transport, indicating efficient generation of the non-linear phonon Hall effect through the ferroelectricity of the SnS monolayer. In circular forced oscillations, a net phonon current also exists along the $x$-direction. However, this phonon flow is independent of the circular polarization of forced oscillations; {\it i.e.} there is no circular dichroic phonon rectification along the $x$-direction (see Supporting Information for details). From the handedness-independence, the phonon transport along the $x$-direction is a tail of the phonon flow shown in Fig. 4(b), which could be attributed to the linear component of the circular forced oscillation.

Recently, it is feasible to excite a selective phonon vibrational mode using terahertz-frequency optical pulses. Successful examples include phonon-driven high-temperature superconductivity \cite{Mankowsky2014}, metal-insulator transition \cite{Rini2007}, an effective magnetic field control \cite{Nova2017}, and ferroelectricity \cite{Nova2019}. With the aid of this state-of-the-art experimental method, the non-linear phonon Hall effect demonstrated in our NEMD simulations can be realized by resonantly exciting a specific phonon mode (\#7) with terahertz laser pulses.

\vspace{2mm}

Here, we presented an efficient way to generate and control phonon BC through ferroelectricity, and simulated non-linear phonon Hall and circular dichroic phonon transport. This concept can be easily applied to other bosonic systems, bringing BC-related non-linear transport of various bosons into a wide range of applications. For example, we can design inversion-broken acoustic lattices, phononic and photonic crystals, and cold atom systems composed of polar molecules~\cite{Xiao2015,Lu2013,Micheli2006}. Moreover, given that the spin Rashba effect greatly expanded the scope of spintronics, the newly introduced phonon Rashba effect has the potential to open the door to phonon research of spintronics applications. By generating an effective momentum-dependent magnetic field coupled with pseudo-angular momentum of phonon, the phonon Rashba effect may realize the {\it electric control of chiral phonons}, such as the phonon angular momentum Hall effect~\cite{Park2020} and phonon Edelstein effect~\cite{PhysRevLett.121.175301}.
\vspace{4mm}

\section{Acknowledgements}
The authors gratefully acknowledge financial support from the National Research Foundation of Korea (NRF) under Grant 2017M3A7B4049273 (J.~I.), the Institute for Basic Science under Grant IBS-R009-D1 (C.~H.~K.), and the National Research Foundation of Korea (NRF) under Grants 2021M3H4A1A03054864, 2019R1A2C1010498 and 2017M3D1A1040833 (H.~J.).\\

\providecommand{\latin}[1]{#1}
\makeatletter
\providecommand{\doi}
  {\begingroup\let\do\@makeother\dospecials
  \catcode`\{=1 \catcode`\}=2 \doi@aux}
\providecommand{\doi@aux}[1]{\endgroup\texttt{#1}}
\makeatother
\providecommand*\mcitethebibliography{\thebibliography}
\csname @ifundefined\endcsname{endmcitethebibliography}
  {\let\endmcitethebibliography\endthebibliography}{}


\end{document}